\documentclass[10pt,twocolumn,letterpaper]{article}

\usepackage{cvpr}              

\usepackage{microtype}

\renewcommand{\paragraph}[1]{\vspace{.5em}\noindent\textbf{#1.}}

\usepackage{soul}
\setuldepth{foobar}

\usepackage{multirow}
\usepackage{arydshln}
\usepackage{listings}
\usepackage[most]{tcolorbox}
\usepackage[T1]{fontenc}
\usepackage{textcomp}

\lstdefinestyle{ieeeprompt}{
  basicstyle=\ttfamily\scriptsize,
  columns=fullflexible,
  breaklines=true,
  breakatwhitespace=true,
  keepspaces=true,
  showstringspaces=false,
  upquote=true,
  aboveskip=0.5\baselineskip,
  belowskip=0.5\baselineskip
}

\newtcblisting{promptbox}[2][]{%
  enhanced,
  arc=1mm, 
  breakable,   
  boxrule=0.4pt,
  colback=white,
  colframe=black!95,
  titlerule=0pt,    
  toptitle=0mm,        
  bottomtitle=0mm,     
  top=0.6mm,           
  left=1.2mm,right=1.2mm,top=0.8mm,bottom=0.8mm,
  listing only,
  listing options={style=ieeeprompt},
  title=\textbf{#2},
  fonttitle=\footnotesize,
  #1
}

\definecolor{cvprblue}{rgb}{0.21,0.49,0.74}
\usepackage[pagebackref,breaklinks,colorlinks,allcolors=cvprblue]{hyperref}

\def\paperID{*****} 
\def\confName{CVPR}
\def\confYear{2026}

\title{Intent-aligned Autonomous Spacecraft Guidance via Reasoning Models}

\author{Yuji Takubo, Simone D'Amico\\
Stanford University \\
496 Lomita Mall, Stanford, CA, 94305\\
{\tt\small \{ytakubo, damicos\}@stanford.edu}
}

\begin{document}
\maketitle

\begin{figure*}[t]
    \centering
    \includegraphics[width=0.86\textwidth]{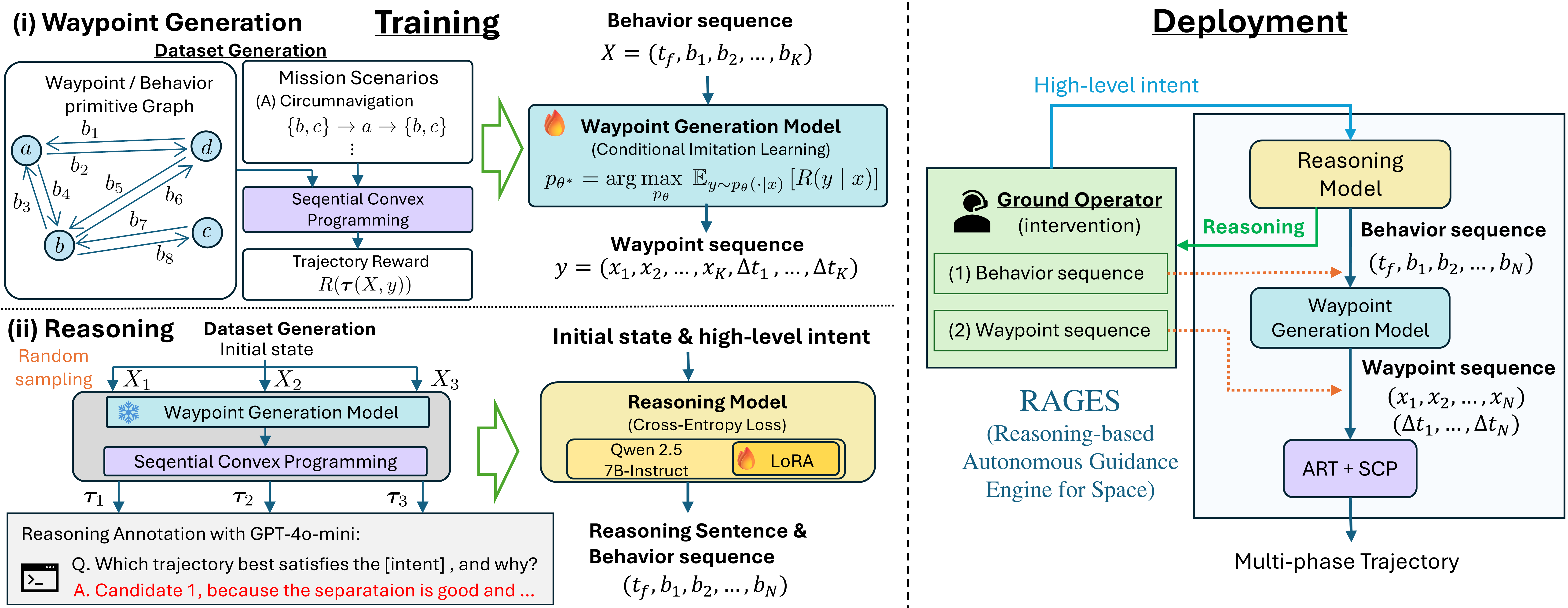}
    \caption{\textbf{Training and deployment of the proposed intent-to-trajectory pipeline.}
    The framework comprises (i) a reasoning model that predicts a behavior sequence from scenario context and high-level intent, (ii) a waypoint generator that produces waypoint constraints, and (iii) an SCP solver that enforces dynamics and safety.
    \emph{Training:} the waypoint generator is trained first using SCP rollouts; the reasoning model is then fine-tuned using the training data generated by evaluating candidate behavior sequences through the frozen waypoint generator and SCP, with automated annotation by a pretrained LLM. 
    \emph{Deployment:} the reasoning model and waypoint generator run online, and SCP produces the final feasible trajectory. The operator can override the generated decision at both the behaviors and the waypoints. }
    \label{fig:rages}
\end{figure*}

\begin{abstract}
Future spacecraft operations require autonomy that can interpret high-level mission intent while preserving safety. 
However, existing trajectory optimization still relies heavily on expert-crafted formulations and does not support intent-conditioned decision-making. 
This paper proposes an intent-aligned spacecraft guidance framework that links high-level reasoning and safe trajectory optimization through explicit intermediate abstractions, based on behavior sequences and waypoint constraints. 
A foundation model first predicts an intent-aligned behavior plan, a waypoint generation model then converts it into waypoint constraints, and the safe trajectory is computed via optimization. 
This decomposition enables scalable supervision without sacrificing safety. 
Numerical experiments in close-proximity operation scenarios demonstrate that the proposed pipeline achieves over 90\% SCP convergence and yields a $1.5\times$ higher rate of generating trajectories that satisfy the top intent-prioritized performance criteria than heuristic decision-making. 
These results support the use of intermediate behavior abstraction as a practical interface between foundation-model reasoning and safety-critical onboard spacecraft autonomy.
\end{abstract}

\vspace{-2em}
\section{Introduction}
\label{sec:intro}

Safe and autonomous guidance and control are essential for future in-space logistics.
During critical phases, spacecraft must transition from ground supervision to goal-driven onboard autonomy, requiring systems that reason over high-level objectives while guaranteeing real-time safety and dynamic feasibility.
Although numerical optimal control methods \cite{malyuta_scp_2022} enable fast nonconvex trajectory generation, they face two key limitations: (i) significant domain expertise is required to translate mission intent into safe, executable trajectories, and (ii) proximity operations still rely heavily on heuristic waypoint-based strategies \cite{damico_phd_2010}.
While effective, these approaches confine strategic reasoning to human design processes or rigid rule-based logic.

Concurrently, foundation models (FMs) have demonstrated general reasoning capabilities and are increasingly adapted for physical AI through multimodal perception-to-action pipelines \cite{lynch2020language, kawaharazuka2025vision, wang2025alpamayo}. 
Nonetheless, an end-to-end reasoning pipeline that ensures the safety and feasibility of the generated trajectory is lacking in space applications. 

This paper proposes an intent-aligned spacecraft guidance framework that translates high-level mission intent into a safe, dynamically feasible, and semantically grounded trajectory through three stages: (i) behavior planning with a reasoning trace, (ii) structured waypoint generation, and (iii) constrained nonconvex trajectory optimization. 
In this hierarchy, behavior sequences provide the key intermediate abstraction between foundation-model reasoning and safe trajectory optimization, while waypoint constraints geometrically ground the selected behavior plan. 
This decomposition enables scalable supervision, systematic incorporation of expert knowledge, and interpretable intervention points at the behavior and waypoint levels. 
As a result, the framework supports progressive autonomy and practical deployment while maintaining operational oversight.

The main contributions of this paper are as follows:
\begin{itemize}
    \item An intent-aligned spacecraft guidance architecture that links foundation model reasoning to safe trajectory generation through behavior and waypoint abstractions.
    \item A staged bootstrapping strategy that constructs scalable supervision for waypoint generation and behavior-level reasoning from optimization-derived trajectory outcomes.
    \item End-to-end validation in a close-proximity inspection scenario, showing over 90\% SCP convergence and improved intent-aligned trajectory generation over heuristics.
\end{itemize}

\vspace{-0.5em}
\section{Related Work}
\label{sec:related_work}

\vspace{-0.5em }
\textbf{Learning-Based Nonconvex Trajectory Generation:}
Sequential convex programming (SCP) \cite{malyuta_scp_2022} has emerged as a leading framework for nonconvex trajectory generation due to its convergence guarantees \cite{bonalli_2019_gusto, mao2016successive, oguri2023successive} and ability to accommodate complex constraints \cite{szmuk2020successive, elango2025continuous}. 
To mitigate sensitivity to initialization, learning-based warm-start strategies have therefore been increasingly explored \cite{Banerjee_2020, kim2022guided, art_ieeeaero24, takubo2024towards, celestini2025generalizable}. 
These approaches leverage neural networks to predict initial trajectories or control sequences that accelerate convergence and improve robustness. 
Recent work extends beyond enhancement and demonstrates the fast generation of multi-objective trade spaces \cite{takubo2026agile}. 
This paper is inspired by prior work on semantic trajectory generation conditioned on text inputs via language encoders, in which safety is enforced via a subsequent SCP \cite{takubo2026semantic}. 
While effective, it relies on text–trajectory pairs constructed from a limited, predefined waypoint set, limiting generalization and scalability.

\textbf{Foundation Models for Robotic Reasoning:} 
Foundation model (FM)–based reasoning has gained traction in robotics as a scalable mechanism for encoding cross-domain knowledge and task priors  \cite{brohan2023saycan, huang2023voxposer, kawaharazuka2025vision, wang2025alpamayo}.
In space applications, recent works explore language-conditioned control and semantic reasoning \cite{carrasco2025large, foutter2025space-llava, jain2026autonomous}, but typically map commands directly to control inputs without enforcing dynamically feasible, constraint-satisfying trajectories. 
High-stakes space operations further require interpretability and operator oversight, which fully end-to-end architectures often obscure.

\vspace{-0.7em}
\section{Methodology}
\label{sec:methodology}

The objective of this paper is to map a high-level mission intent, specified by the operator and encoded with performance priorities, to a dynamically feasible trajectory.
To enable this \textit{intent-to-trajectory} pipeline, a tri-level decision-making architecture is proposed (cf. Fig.~\ref{fig:rages}), which comprises:
(i) a FM that interprets the scenario context and infers a mid-level behavior sequence;
(ii) a waypoint generation model that converts this sequence into structured waypoint constraints; and
(iii) a nonconvex trajectory generation module that synthesizes a feasible trajectory.
The hierarchy is enabled by the intermediate representations: behavior primitives and waypoint constraints. 
Because high-level objectives are inherently under-constrained and admit many feasible trajectories, direct intent-to-trajectory learning can be unstable and data-inefficient. 
Decomposing intent into structured behavior sequences increases decision specificity and improves the robustness of trajectory synthesis, while providing multiple checkpoints for operator intervention. 

Training follows a staged bootstrapping process aligned with the hierarchy. 
The waypoint generator is first trained using the trajectories solved via SCP and labeled with performance metrics. 
The reasoning model is then trained with data that couples high-level intent and trajectory-level quality via the trained waypoint model. 
This sequential procedure propagates optimization-level supervision to higher-level semantic reasoning. 
The subsequent sections introduce the modules in reverse inference order: trajectory generation via SCP, waypoint generation, and finally the reasoning model.

\vspace{-0.3em}
\subsection{Trajectory generation via SCP}

The foundation of the proposed architecture is the nonconvex trajectory generation problem $\mathcal{P}$, which is shown as: 
\begin{subequations} \label{eq:ocp}
\footnotesize
\begin{alignat}{2}
    \min_{\{{x}_j, {u}_j\}_{j=1}^{N}} \quad & 
    \sum_{j=1}^{N} \|u_j\|_2 \label{eq:ocp_obj}\\
    \text{subject to} \quad 
    & {x}_{j+1} = \Phi (t_{j+1}, t_j) ({x}_{j} + \Gamma_j u_j), \  
    && \forall j \in \mathbb{Z}_{1:N-1}, \label{eq:ocp_con_dyn} \\
    & x_1 = \hat{x}_{0}, \quad x_{j_k} = \tilde{x}_k, && \forall k \in \mathbb{Z}_{1:K},  \label{eq:ocp_con_wyp} \\ 
    & g(x_j, u_j) \leq 0, && \forall j \in \mathbb{Z}_{1:N-1}. \label{eq:ocp_con_safety}
\end{alignat}
\end{subequations}
The state vector ${x} \in \mathbb{R}^6$ is in quasi-nonsingular Relative Orbital Elements (qnsROE)~\cite{damico_phd_2010}, and the objective minimizes total fuel expenditure.
Eq.~\eqref{eq:ocp_con_dyn} is discrete-time dynamics with impulsive inputs $u \in \mathbb{R}^3$ applied in the Radial--Tangential--Normal (RTN) frame; Eq.~\eqref{eq:ocp_con_wyp} is the waypoint-passage constraints parameterized by $(\tilde{x}_k, j_k)$ and the initial state $\hat{x}_0$; and Eq.~\eqref{eq:ocp_con_safety} is a nonconvex safety constraint. 
In particular, chance-constrained passive safety~\cite{guffanti_jgcd_2023, takubo2024towards} is enforced to guarantee separation from the ellipsoidal keep-out zone (KOZ) with a probabilistic margin, even under sudden loss of control authority.
This problem is solved via SCP~\cite{malyuta_scp_2022}.

\vspace{-0.5em}
\subsection{Waypoint generation model}

The waypoint generation model predicts a waypoint placement policy conditioned on an intermediate sequence of \emph{behavior primitives}.
Its output is $y = \bigl(\{\tilde{x}_k, d_k\}_{k=1}^{K}\bigr)$, where $\tilde{x}_k$ denotes the $k$-th waypoint state and $d_k$ is the corresponding transfer duration in discrete steps.
With a fixed time grid of size $\Delta t$, the total maneuver time is $t_f = \sum_{k=1}^{K} d_k\,\Delta t.$
The model is conditioned on $X = \bigl(\hat{x}_0,\, t_f,\, \{b_k\}_{k=1}^{K},\, \boldsymbol{z}\bigr)$, where $\hat{x}_0$ is the initial belief state, $t_f$ is the maneuver horizon, $\{b_k\}_{k=1}^{K}$ is the behavior-primitive sequence, and $\boldsymbol{z} = \bigl(\oe(t_0),\, r_{\mathrm{KOZ}},\, \beta\bigr)$ collects scenario parameters: target orbital elements, the keep-out-zone radius, and a scalar uncertainty parameter, respectively.

Waypoint generation is intended to approximate the trade-offs made by mission designers to maximize overall trajectory quality. 
As manual annotation of high-quality waypoint strategies is not scalable, a proxy reward is defined based on the solution of a trajectory optimization problem $\mathcal{P}(X,y)$. 
Given input $X$ and waypoint configuration $y$, solving $\mathcal{P}(X,y)$ via SCP yields a feasible trajectory $\boldsymbol{\tau}^*_{\mathcal{P}(X,y)} = \{x_j^*, u_j^*\}_{j=1}^{N}$, from which a scalar reward $R$ is computed.
In particular, this paper defines $R$ as a weighted sum of the control cost and trajectory-wide observation quality. 
By restricting the design space to waypoint placement, the optimization of $R$ becomes tractable.

The waypoint generator is modeled as a conditional Gaussian distribution whose parameters are predicted by a multi-layer perceptron (MLP). 
It is trained via weighted maximum likelihood estimate, where each sample is weighted according to the reward of the corresponding SCP solution.

A dataset of $(X, y, R)$ tuples is constructed via structured scenario generation.
Each behavior primitive corresponds to a transfer between relative orbital domains, inducing a directed graph in the qnsROE space: nodes represent admissible domains, and edges encode behavior-induced transitions.
Three representative on-orbit inspection scenarios (\emph{Circumnavigation}, \emph{Flyby}, and \emph{Ducking}) define distinct connectivity structures over this graph, each associated with admissible transfer-time windows.
Behavior sequences and corresponding durations are sampled under these structural constraints.
This rule-based curation promotes scenario diversity while excluding semantically inconsistent transitions.

\vspace{-0.5em}
\subsection{Reasoning model}

The reasoning model predicts a dynamically feasible behavior sequence consistent with mission intent. 
It maps the input $\zeta = [\hat{x}_0,\, \boldsymbol{z},\, \text{intent}]$ to the output $\eta = [\text{reasoning},\, t_f,\, b_1,\ldots,b_K]$, which includes a behavior sequence, the total transfer time, and a concise reasoning trace. 
In this work, \textit{intent} is represented as an ordered priority set,
$\text{priority}=\{\text{fuel},\ \text{time},\ \text{observation},\ \text{safety margin}\}$, and the reasoning trace is a single sentence that references one or two top-priority metrics. 
This structured interface enables scalable dataset construction. 
Importantly, at deployment, the reasoning model does not observe downstream waypoint- or trajectory-level metrics; it must infer a behavior sequence solely from the scenario context and the intent specification.

To construct paired data $(\zeta,\eta)$, an annotation pipeline converts optimization-derived performance comparisons into supervision for behavior-level decision-making (cf. Fig.~\ref{fig:rages}). 
For fixed $(\hat{x}_0,\boldsymbol{z},\text{intent})$, $M=4$ candidate mid-level decisions are sampled, each specifying a valid behavior-primitive sequence and total time $t_f$.
Each candidate is passed through the frozen waypoint generator and the SCP that solves Eq.~\ref{eq:ocp}, treated as a black-box simulator.
This yields a feasible trajectory and four quantitative metrics corresponding to the intent priority list. 
A pretrained LLM is used during dataset construction to convert structured candidate comparisons into a selected behavior plan and concise rationale consistent with the ordered intent priorities.
This produces structured targets that contain both the selected behavior decision and its reasoning trace; trajectory-level metrics are used only for supervision during training.

The reasoning model is trained via supervised fine-tuning of a pretrained LLM (Qwen2.5-7B-Instruct). 
A token-wise cross-entropy loss is applied to the autoregressive output sequence, covering both behavior tokens and reasoning tokens, with a weighting factor controlling their relative contributions. 
Low-Rank Adaptation (LoRA) \cite{hu2022lora} is employed for parameter-efficient fine-tuning by introducing trainable low-rank updates to frozen transformer layers.

\vspace{-0.7em}
\section{Results}

\begin{figure}[t!]
    \centering
    \includegraphics[width=1.05\linewidth]{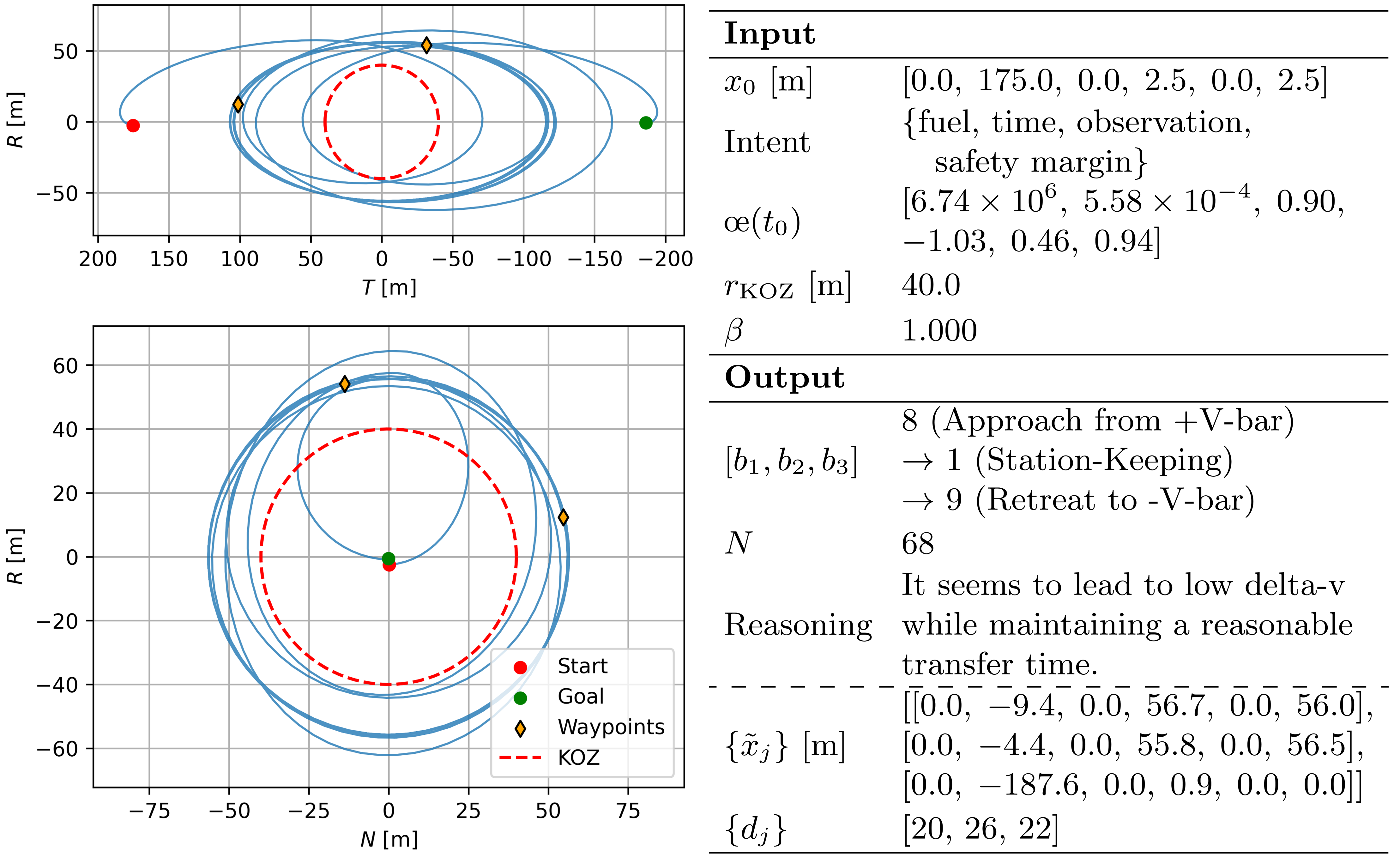}
    \caption{Representative Trajectory Generation.}
    \label{fig:rep_soln}
\end{figure}

Figure~\ref{fig:rep_soln} summarizes representative inputs and outputs, including the inferred intermediate decisions and the resulting multi-phase RTN-frame trajectory.
The solution generates a behavior sequence aligned with the highest-priority intent metric, followed by waypoint constraint construction and SCP-based trajectory optimization to enforce passive safety.

\vspace{-0.6em}
\subsection{Performance of waypoint generation}

\begin{table}[t] 
\centering
\scriptsize
\begin{tabular}{llccc}
\hline
 &  & \multicolumn{2}{c}{Neural Module} &  \\
\cline{3-4}
 &  & Unweighted & Weighted & Heuristic \\
\hline
\multicolumn{2}{l}{SCP Success [\%]} & \textbf{90.8} & 90.6 & 72.2 \\
\multicolumn{2}{l}{Reward mean / std.} & -2.99 $\pm$ 0.95 & \textbf{-2.91 $\pm$ 0.96} & -3.28 $\pm$ 1.14 \\
\hline
\multirow{3}{*}{\shortstack{max. \\ waypoint \\ error [\%]}}
 & 0 m  & \textbf{62.4} & 56.4 & 100 \\
 & $<$ 5 m  & 83.8 & \textbf{81.6} & 100 \\
 & $<$ 10 m & 89.4 & \textbf{98.8} & 100 \\
\hline
\end{tabular}
\caption{Statistical performance of the waypoint generation. }
\label{tab:wyp_gen_summary}
\end{table}

The waypoint generation model is first evaluated to validate the effectiveness of the bootstrapped training procedure. 
A total of 50,000 $(X,y,R)$ samples are generated with a 90–10 train–validation split, and performance is assessed on 500 independently sampled test cases using deterministic mean predictions. 
Three approaches are compared: reward-weighted/unweighted imitation learning and a heuristic waypoint placement method based on relative orbit geometry.
In the heuristic baseline, waypoints are randomly sampled within the admissible domains defined by the waypoint–behavior primitive graph, subject to predefined geometric constraints. 
This guarantees that all sampled waypoints remain within the targeted domain.

As shown in Table~\ref{tab:wyp_gen_summary}, both neural models improve the downstream SCP success rate to approximately 90\%, compared to the 72\% in the heuristic baseline. 
The reward-weighted model achieves the highest mean reward with comparable variance, indicating modest but consistent gains in trajectory quality when reward information is incorporated during training.
Furthermore, roughly 60\% of generated waypoint sequences lie strictly within the prescribed geometric waypoint domains. 
However, over 98\% of waypoints fall within a 10\,m tolerance of their designated domains for the weighted imitation learning, demonstrating strong structural alignment even when exact domain membership is not satisfied.

\vspace{-0.5em}
\subsection{End-to-end evaluation}

\begin{table}[t]
\centering
\scriptsize
\begin{tabular}{lcccc}
\hline
 Reasoning &  Heur. & Heur. &  Neural &  Neural \\
Waypoint Gen.  & Heur. & Neural & Heur. & Neural \\
 \hline 
Behavior-Seq Feasibility [\%] & \multicolumn{2}{c}{--} & \multicolumn{2}{c}{100} \\
SCP Success [\%] & 75.2 & \textbf{91.6} &  65.2  & 90.2 \\
\hdashline
Intent-Reasoning match (2/2) [\%] & \multicolumn{2}{c}{--}  & \multicolumn{2}{c}{52.4} \\
Intent-Reasoning match: (1/2) [\%] & \multicolumn{2}{c}{--}  & \multicolumn{2}{c}{47.6}  \\
Intent-Reasoning match: (0/2) [\%] & \multicolumn{2}{c}{--}  & \multicolumn{2}{c}{0.0}  \\
Reason-Traj. Dual-Metric Win [\%] & 5.0 & 12.4  & 3.8 & \textbf{20.4} \\
Reason-Traj. Single-Metric Win [\%] & 38.6 &  41.4  & 25.4& \textbf{45.6} \\
Intent-Traj. Dual-Metric Win [\%] & 7.4 & 13.0 & 3.6  & \textbf{18.0} \\
Intent-Traj. Single-Metric Win [\%] & 35.6 &  42.2 & 24.4 &\textbf{44.0} \\
\hline
\end{tabular}
\caption{End-to-end performance analysis with the consistency between the high-level intent, reasoning, and final trajectory, across four decision-making pipelines based on heuristics (Heur.) or trained neural modules (Neural).}
\label{tab:rages_e2e}
\end{table}

Finally, an end-to-end statistical evaluation of the proposed hierarchical pipeline is conducted across four configurations, defined by combinations of heuristic and trained neural modules for behavior-sequence reasoning and waypoint generation. 
For heuristic reasoning, a feasible behavior sequence initiating from the waypoint domain to which the initial state belongs is randomly sampled. 
Three categories of metrics are considered: (i) behavior-sequence feasibility, assessing whether the reasoning module produces an admissible behavior sequence; (ii) success rate in the downstream SCP convergence; and (iii) semantic consistency.

Semantic consistency is quantified by measuring alignment between high-level intent, reasoning traces, and resulting trajectory performance.
For each scenario, up to two prioritized metrics are extracted from the reasoning trace using GPT-4o-mini and compared against the top two priorities specified in the high-level intent to quantify intent–reasoning alignment. 
The final converged trajectory is then compared with the three alternatives generated from the same scenario input $\zeta$. 
A \textit{dual (single)-metric win} is recorded when two (one) out of two prioritized metrics outperform those of the other candidates. 
During evaluation, both the reasoning and waypoint generation models operate deterministically to ensure reproducibility.

Table~\ref{tab:rages_e2e} summarizes the results over 500 test scenarios for all combinations of heuristic and trained (neural) reasoning and waypoint modules. 
The results show that the trained reasoning model achieves 100\% feasibility in generating feasible behavior sequences. 
Also, the SCP convergence rate is largely governed by the waypoint generation, where the trained model substantially improves the success rate under both reasoning settings.

In terms of semantic alignment, more than 50\% of reasoning traces match both top-priority intent metrics, and all match at least one. 
The fully trained configuration also achieves the highest dual-metric win rates for both intent–trajectory and reasoning–trajectory alignment, yielding approximately a $1.5\times$ relative improvement (18.0\% and 20.4\%) over heuristic reasoning with the same trained waypoint generator (13.0\% and 12.4\%), which demonstrates the effectiveness of introducing a neural module for generating mid-level behavior sequences. 

Overall, the results show that the hierarchical architecture successfully propagates high-level intent through intermediate behavior representations to trajectory-level performance. 
In particular, training the waypoint generator improves SCP convergence, and bootstrapping the reasoning model with trajectories from the trained waypoint model enhances semantic consistency, enabling the intent-aligned safe trajectory generation.

\vspace{-0.7em}
\section{Conclusion and Future Work}

This paper presents an intent-aligned autonomous spacecraft guidance framework that links reasoning and safe trajectory generation through behavior-level abstraction and waypoint generation.  
Future work will incorporate learning-based warm-start strategies and a more rigorous analysis of the generalization properties of the elicited reasoning capability.

\vspace{-0.7em}
\section{Acknowledgement}

Yuji Takubo thanks the Ezoe Memorial Recruit Foundation for the financial support. 
The authors thank Daniele Gammelli and Maro Pavone for helpful discussions. 

{
    \small
    \bibliographystyle{ieeenat_fullname}
    \bibliography{main}
}

\clearpage
\setcounter{page}{1}
\maketitlesupplementary

\section{Nonconvex Trajectory Generation}
\label{sec:supp_scp}

The nonconvex trajectory optimization in Eq.~\eqref{eq:ocp} solves for an optimal open-loop trajectory that is robust to the uncertainty. 
Discrete-time dynamics of the qnsROE space are expressed in Eq.~\eqref{eq:ocp_con_dyn}. 
In particular, $\Phi(t_{j+1}, t_j) \in \mathbb{R}^{6\times6}$ is the state transition matrix of qnsROE with the secular $J_2$ effect~\cite{koenig2017new} and $\Gamma_j = \Gamma(t_j) \in \mathbb{R}^{6\times3}$ is the control input matrix~\cite{damico_phd_2010}. 

Furthermore, the nonconvex safety constraint in Eq.~\eqref{eq:ocp_con_safety} ensures the spacecraft is outside of the ellipsoidal safety domains around the target not only for the entire duration of the controlled trajectory $t\in [0, t_f]$ but also for a predefined duration of the drift trajectories $\tau \in [0, \tau^s]$ after the abrupt loss of control. 
Formally, this is expressed as the chance-constraint form as \cite{takubo2026agile}:
\begin{subequations}
\small
\begin{align}
    & \Pr( {x}_{j0}^\top S_{ki} {x}_{j0}  \geq 1) \geq 1-\Delta_j, \label{eq:con_ps_chance} \\
    & \Phi_{ji} = \Phi(t_j + \tau_i, t_j), \ \ \Psi_{ji} = \Psi(t_j+\tau_i), \\
    & S_{ki} = (\Psi_{ji} \Phi_{ji})^\top P \Psi_{ji} \Phi_{ji}, \\
    & {x}_{j0} = {x}_j+\Gamma_j{u}_j,
\end{align}
\end{subequations}
where $\Psi (t) \in \mathbb{R}^{6\times6}$ is the first-order mapping between the qnsROE and the relative position and velocity vector resolved in the RTN frame \cite{damico_phd_2010}, $\Delta_j$ is the prescribed risk factor, and $P \in \mathbb{R}^{6\times6}$ is a diagonal matrix with its components corresponding to the semi-axes of the KOZ in the RTN frame.
Under the assumption of Gaussian uncertainty, this chance-constraint is rewritten to a deterministic form as follows:
\begin{subequations} 
\label{eq:con_ps_det}
\footnotesize
\begin{align}
    & 1 - {x}_{j0}^\top S_{ki} {x}_{k0} + q(\Delta_k) \sqrt{\boldsymbol{g}_{ki}^\top \Sigma_{ji} \boldsymbol{g}_{ji}}  \leq 0 \\
    & \boldsymbol{g}_{ji} = -2 \Phi_{ji}^{-\top}(S_{ji} {x}_{j0}), \\
    & \Sigma_{j(i+1)} = \Phi_{ji} \Sigma_{ji} \Phi_{ji}^\top + Q_{ji}, \ \  \Phi_{ji} = \Phi(t_j+\tau_{i+1}, t_j+\tau_{i}), \\
    & \Sigma_{j0} = \Sigma_{\text{nav}}({x}_j) + \Gamma_j \Sigma_{\text{exe}} ({u}_j) \Gamma_j^\top, \label{eq:con_cov_ini}
\end{align}
\end{subequations}
where $q(\cdot)$ is the inverse of the normal cumulative distribution function; $Q_{ji}$ is the constant discrete-time process noise covariance for unmodeled acceleration; $\Sigma_{\text{nav}}({x}_j)$ is the navigation uncertainty, and $\Sigma_{\text{exe}}(u_j)$ is the control execusion error modeled by the Gates model  \cite{gates1963simplified}.
The navigation covariance in qnsROE is modeled as a function of the range between the spacecraft and the target $\rho({x}_j)$ as:
\begin{subequations}
\begin{align} \label{eq:nav_cov} \footnotesize
    & \Sigma_{\text{nav}}({x}_j) = \rho({x}_j) \boldsymbol{s} \boldsymbol{s}^\top, \\
    & \boldsymbol{s} = \beta \cdot 10^{-3} \cdot [0.1, 4.0, 2.0, 2.0, 2.0, 2.0] \text{[m]},
\end{align}
\end{subequations}
where scalar variable $\beta$ defines the navigation accuracy.

\section{Waypoint Generation Model}
\label{sec:wyp_gen}

\subsection{Dataset generation with waypoint/behavior primitive graph}

The generated dataset is constructed from five domains in the qnsROE space, summarized in Table~\ref{tab:roe_domain}.
Note that $\delta a=\delta e_x = \delta i_x =0$ is set so that each waypoint does not have an along-track drift, having no first-order $J_2$-perturbation, and having the maximal RN-plane separation by leveraging the E/I-vector separation \cite{damico_phd_2010}. 
This reduces the sampling domain from originally a six-dimensional space to a two-dimensional space. 
Table~\ref{tab:behav_prim} defines the behavior primitives that enable transitions between these qnsROE domains, while Table~\ref{tab:campaign} specifies the campaign-level qnsROE sequences.
Together, these three tables define the waypoint–behavior primitive graph from which waypoint sequences are sampled during dataset generation.
\begin{table}[ht!] 
\centering
\caption{Canonical relative orbit domains expressed in the qnsROE space.}
\label{tab:roe_domain}
\small
\begin{tabular}{lcc}
\hline
\textbf{Name} & $\delta \lambda$ & $\delta e_y = \delta i_y$ \\
\hline
(a) Central, passively safe orbit     
& $[-5,\,5]$          
& $[30,\,70]$ \\
(b) +V-bar, passively safe orbit      
& $[100,\,250]$                
& $[30,\,70]$ \\
(c) +V-bar axis                       
& $[100,\,250]$                
& $[-5,\,5]$ \\
(d) -V-bar, passively safe orbit      
& $[-250,\,-100]$                
& $[30,\,70]$ \\
(e) -V-bar axis                       
& $[-250,\,-100]$                
& $[-5,\,5]$ \\
\hline
\end{tabular}
\end{table}

\begin{table}[ht]
\small
\centering
\caption{Behavior-mode transitions. } \label{tab:behav_prim}
\begin{tabular}{ll}
\hline
\textbf{Behavior Mode} & \textbf{Transition} \\
\hline
(1) Station-Keeping & \multicolumn{1}{l}{Self-loop at all nodes: (a)–(e)} \\
(2) Drift +V-dir.                 & (d)$\rightarrow$(a), (a)$\rightarrow$(b), (d)$\rightarrow$(b) \\
(3) Drift -V-dir.                 & (b)$\rightarrow$(a), (a)$\rightarrow$(d), (b)$\rightarrow$(d) \\
(4) Expand R/N separation         & (e)$\rightarrow$(d), (c)$\rightarrow$(b) \\
(5) Shrink R/N separation         & (b)$\rightarrow$(c), (d)$\rightarrow$(e) \\
(6) Approach from -V-bar          & (e)$\rightarrow$(a) \\
(7) Retreat to +V-bar             & (a)$\rightarrow$(c) \\
(8) Approach from +V-bar          & (c)$\rightarrow$(a) \\
(9) Retreat to -V-bar             & (a)$\rightarrow$(e) \\
(10) Ducking (fast drift) +V-dir.  & (e)$\rightarrow$(c) \\
(11) Ducking (fast drift) -V-dir.  & (c)$\rightarrow$(e) \\
\hline
\end{tabular}
\end{table}

\begin{table}[ht]
\centering \footnotesize 
\caption{Admissible orbital-region sequences by mission type. The symbol ``$\cdot$'' denotes no transfer (no-op) and is not counted as a region transition.} 
\label{tab:campaign}
\begin{tabular}{lccc}
\hline 
\textbf{Mission Type} & \textbf{Phase 1} & \textbf{Phase 2} & \textbf{Phase 3} \\
\hline 
(A) Circumnav.
& (b$|$c$|$d$|$e)$\rightarrow$(a)
& (a)$\rightarrow$(a)
& (a)$\rightarrow$(b$|$c$|$d$|$e)
\\
\hline
(B) Flyby 
& (c)$\rightarrow$(b) or (b)$\rightarrow$$\cdot$
& (b)$\rightarrow$(d)
& (d)$\rightarrow$(e) or (d)$\rightarrow$$\cdot$
\\
& (e)$\rightarrow$(d) or (d)$\rightarrow$$\cdot$
& (d)$\rightarrow$(b)
& (b)$\rightarrow$(c) or (b)$\rightarrow$$\cdot$
\\
\hline
(C) Ducking 
& (b)$\rightarrow$(c) or (c)$\rightarrow$$\cdot$
& (c)$\rightarrow$(e)
& (e)$\rightarrow$(d) or (c)$\rightarrow$$\cdot$
\\
& (d)$\rightarrow$(e) or (e)$\rightarrow$$\cdot$
& (e)$\rightarrow$(c)
& (c)$\rightarrow$(b) or (c)$\rightarrow$$\cdot$
\\
\hline
\end{tabular}
\end{table}

In addition to the construction of the waypoint-behavior primitive graph, Table~\ref{tab:sim_params} defines the parameter ranges taken for the randomization of the input $X$.

\begin{table}[ht]
\centering
\small
\caption{Simulation and model parameters.}
\label{tab:sim_params}
\begin{tabular}{ll}
\hline
\textbf{parameter} & \textbf{value} \\
\hline
$\Delta t$ [s] & 900 \\
$N_\text{max}$ & 100 ($\sim t_f \leq $ 15 hr.) \\
$r_{\mathrm{KOZ}}$ [m] & $\{20.0,\;30.0,\;40.0\}$ \\
$\beta$ & $\{0.75,\;1.0,\;1.25,\;1.5,\;2.0\}$ \\
\hline 
\multicolumn{2}{l}{$\oe(t_0) = [a,\;e,\;i,\;\Omega,\;\omega,\;M]$} \\
$a$ [m] & $6.73814 \times 10^{6}$ \\
$e$ & $5.581 \times 10^{-4}$ \\
$i$ [$^\circ$] & $51.64$ \\
$\Omega$ [$^\circ$] & $301.04$ \\
$\omega$ [$^\circ$] & $26.18$ \\
$M$ [rad] & $\mathrm{linspace}(0,\,2\pi,\,20)$ \\
\hline
\end{tabular}
\end{table}

\subsection{Reward model}
The reward model used in the waypoint generation model is defined distinctly from the simple fuel cost in Eq.~\eqref{eq:ocp_obj} as follows: 
\begin{align} 
  R(y;X) = - \lambda R_{c}(\boldsymbol{\tau}^*_{\mathcal{P}(X, y)}) + R_{o}(\boldsymbol{\tau}^*_{\mathcal{P}(X,y)}),
\end{align}
where $\boldsymbol{\tau}^*_{\mathcal{P}(X,y)} = \{x_j^*, u_j^*\}_{j=1}^{N}$ denotes a minimum-fuel solution of a nonconvex trajectory generation problem $\mathcal{P}(X,y)$; $R_c$ denotes the control cost, $R_o$ denotes the observational reward associated with the trajectory, and $\lambda$ denotes the weighting hyperparameter. 
Note that, when desired, this can be further extended and set $\boldsymbol{\tau}^*_{\mathcal{P}(X,y)} = \{x_j^*, u_j^*\}_{j=1}^{N}$ to be a high-fidelity trajectory, where the navigation and control error based on Monte-Carlo simulation (e.g., $\Delta V_{99}$) are part of the reward function. 

The control cost and the observational reward are defined as: 
{\small \begin{align} 
R_c & = \sum_{j=1}^{N} \|u_j\|_2, \\ 
R_o & = -\frac{1}{N}\sum_{j=\min\{j: \rho (x_j) \le r_{\mathrm{KOZ}}+\Delta r\}}^{\max\{j: \rho (x_j) \le r_{\mathrm{KOZ}}+\Delta r\}} \rho (x_j).
\end{align}}
This construction selects the contiguous time interval spanning the first and last epochs at which the ego spacecraft enters the observation region of radius $r_{\mathrm{KOZ}} + \Delta r$.
Within this interval, the observational reward increases as the spacecraft remains closer to the target. The cost is then normalized by the total number of timesteps $N$.

\subsection{Training}

In order to bias the policy toward higher-reward samples, a weighted imitation learning objective is adopted as:
\begin{subequations}
\small
\begin{align}
\mathcal{L}(\theta)
&= - \sum_{i=1}^{B} w_i \, \log p_{\theta}\!\left( y^{(i)} \mid X^{(i)} \right), \\
w_i
&= \frac{R_i - R_{\min}}
{\sum_{j=1}^{B} \left( R_j - R_{\min} \right)} .
\end{align}
\end{subequations}
where $w_i\geq 0$ denotes the normalized weight assigned to the $i$-th sample in a batch of size $B$.
The rewards are shifted by $R_\text{min}$ to ensure nonnegativity before normalization.

\section{Reasoning Model}

\subsection{Dataset generation}

First, four quantitative evaluation metrics corresponding to the high-level intent components 
$\{\text{fuel},\, \text{time},\, \text{observation},\, \text{safety margin}\}$ are introduced as follows:
\begin{itemize}
    \item \textbf{Fuel:} $R_c$
    \item \textbf{Time:} $N = \sum_{k=1}^{K} d_k$
    \item \textbf{Observation quality:} $R_o$
    \item \textbf{Safety margin:} $\displaystyle \min_{j} \rho(x_j) - r_{\mathrm{KOZ}}$
\end{itemize}
Higher values of the observation metric $R_o$ and the safety margin are preferred, 
whereas lower values of fuel $R_c$ and total time $N$ are desirable.

\subsection{Supervised Fine-tuning}

The reasoning model is trained based on supervised fine-tuning of a transformer-based pretrained LLM (Qwen2.5-7B-Instruct). 
To reduce memory and computational overhead, the base model is loaded in 4-bit quantized form, and LoRA layers are applied to the attention projection matrices of the self-attention layers. 
Training is performed on structured reasoning traces containing a behavior sequence and the total transfer time, formatted as JSON. 
The model parameters are optimized by minimizing the token-level cross-entropy loss over the serialized output sequence.

\section{Supplementary Results}

\subsection{Waypoint generation}

\begin{figure}[t!]
    \centering
    \includegraphics[width=0.97\linewidth]{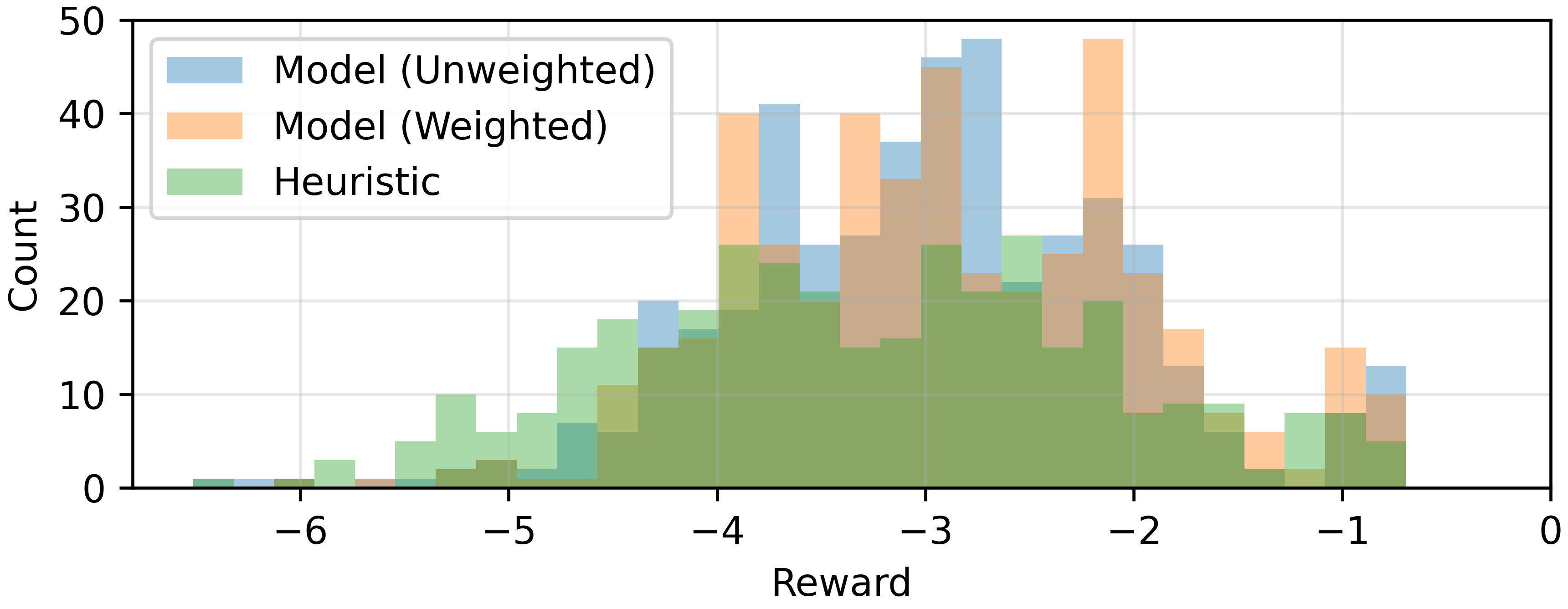}
    \caption{Distribution of the reward function $R(y; X)$ across different waypoint generation models.}
    \label{fig:hist_reward}
\end{figure}

Fig.~\ref{fig:hist_reward} presents the histogram of rewards obtained from the 500 test cases, corresponding to the summary statistics reported in Table~\ref{tab:wyp_gen_summary}.
The superior reward distribution achieved by the imitation learning–based methods (both weighted and unweighted) relative to the heuristic baseline is clearly evident.

\section{LLM Prompts}

\subsection{Annotation of reasoning traces from tabularized metrics}

\begin{promptbox}{Generation of behavior sequence with reasoning (GPT-4o-mini)}
System:
You're an expert spacecraft operator for rendezvous missions.
You select one trajectory candidate from metric tables.
Follow the priority order (lexicographic), not weighted sum.
Output only valid JSON.
For one_line_reason, use probabilistic wording to describe why the selected candidate seems favorable based on its metrics and the intent priority.
Avoid absolute superlatives or explicit comparisons, as the candidates may have tradeoffs and there are no guarantees.
Focus on the strengths of the chosen candidate in relation to the mission intent, without directly stating it is the best or comparing it to others.

User:
Priority order: <intent_1> > <intent_2> > <intent_3> > <intent_4>
Metrics: <metric_1>, <metric_2>, <metric_3>, <metric_4>
Rules:
- Lower is better for fuel_dv and time_sec.
- Higher is better for obs and safety_margin.
- All candidates are already safe; safety_margin is a metric of conservatism.

Candidates CSV:
id,policy,fuel_dv,time_sec,obs,safety_margin
<...rows...>

Return JSON with keys: {"best_candidate_id": <int>, "one_line_reason": "<short sentence>"}

Reasoning style constraints:
- one_line_reason must be exactly one short sentence.
- Do not mention candidate IDs or names.
- Avoid comparative/superlative words: lower, higher, lowest, highest, better, best, worse, worst, more, less.
- Avoid ranking symbols or explicit comparisons: >, <, >=, <=, versus, than.
- Prefer probabilistic phrasing like: it seems to lead to low delta-v, is expected to keep transfer time short, has a high chance of supporting observation.
\end{promptbox}

\subsection{Generation of behavior sequence with a reasoning trace}

\begin{promptbox}{Generation of behavior sequence with reasoning (Qwen2.5-7B-Instruct)}
System:
You are an assistant that selects trajectory-level decisions from structured mission context.
Follow intent priority and constraints, reason briefly, and return a strict JSON answer.

User:
You're an expert spacecraft operator for rendezvous missions.
Task: choose (b_seq, tf) based on mission context and intent priority.
Then provide one-line reasoning and justification.

x0_roe_m = <x0>
r_koz = <r_koz>
beta = <beta>
intent_priority = <intent_priority>

Return JSON with keys: reasoning, tf, b_seq.
\end{promptbox}

\begin{promptbox}{Structured output from the reasoning model (JSON format)}
<|think|>
<reasoning text>
<|answer|>
{"reasoning":"...","tf":...,"b_seq":[...]}
<|end|>
\end{promptbox}

\subsection{Extraction of two top performance metrics from the generated reasoning trace }

\begin{promptbox}{LLM prompt for metric extraction (GPT-4o-mini)}
System: 
Extract metric names exactly from the allowed list.

User:
From the reasoning sentence, select up to two metrics mentioned in order of appearance.
Allowed metrics: fuel_dv, transfer_time_sec, observation_score, safety_margin_m.
Some words to check for each metric:
- fuel_dv: fuel, delta-v, control cost
- transfer_time_sec: time, transfer time, tof
- observation_score: observation
- safety_margin_m: safety, safety margin, clearance
Return strict JSON only: {"focused_metrics": ["..."]}
Reasoning sentence:
<REASONING_SENTENCE>
\end{promptbox}

\end{document}